\documentclass[aps,prl,twocolumn,showpacs,groupedaddress]{revtex4}  
\usepackage{graphicx}  
\usepackage{dcolumn}   
\usepackage{bm}        
\usepackage{amssymb}   

\newcommand{\ttb}{$t\bar{t}$}

\newcommand{\ppb}{$p\bar{p}$}

\newcommand{\invpb}{pb$^{-1}$}
\newcommand{\invfb}{fb$^{-1}$}

\newcommand{\met}{\mbox{\ensuremath{\,\slash\kern-.7emE_{T}}}}

\newcommand{\mht}{\mbox{\ensuremath{\,\slash\kern-.7emH_{T}}}}

\def\lsim{\mathrel{\rlap{\lower4pt\hbox{$\sim$}}
    \raise1pt\hbox{$<$}}}                

\newcommand{\HT}{H_T}

\newcommand{\etadet}{\vert\eta_{\mathrm{det}}\vert}

\newcommand{\xo}{\tilde{{\chi}}^0_1}

\begin{document}


\hspace{5.2in} \mbox{Fermilab-Pub-08/271-E}

\title{Search for scalar leptoquarks and T-odd quarks in the acoplanar jet topology\\
using 2.5\,fb$^{\bm{-1}}$ of $\bm{p\bar{p}}$ collision data at $\bm{\sqrt{s}=}$1.96\,TeV
}
%
\author{V.M.~Abazov$^{36}$}
\author{B.~Abbott$^{75}$}
\author{M.~Abolins$^{65}$}
\author{B.S.~Acharya$^{29}$}
\author{M.~Adams$^{51}$}
\author{T.~Adams$^{49}$}
\author{E.~Aguilo$^{6}$}
\author{M.~Ahsan$^{59}$}
\author{G.D.~Alexeev$^{36}$}
\author{G.~Alkhazov$^{40}$}
% ...

\affiliation{\vspace{0.1 in}(The D\O\ Collaboration)\vspace{0.1 in}}
% ...
\affiliation{$^{6}$University of Alberta, Edmonton, Alberta, Canada,
                Simon Fraser University, Burnaby, British Columbia, Canada,
                York University, Toronto, Ontario, Canada, and
                McGill University, Montreal, Quebec, Canada}
% ...
\affiliation{$^{29}$Tata Institute of Fundamental Research, Mumbai, India}
% ...
\affiliation{$^{36}$Joint Institute for Nuclear Research, Dubna, Russia}
% ...
\affiliation{$^{40}$Petersburg Nuclear Physics Institute,
                St. Petersburg, Russia}
% ...
\affiliation{$^{49}$Florida State University, Tallahassee, Florida 32306, USA}
% ...
\affiliation{$^{51}$University of Illinois at Chicago,
                Chicago, Illinois 60607, USA}
% ...
\affiliation{$^{59}$Kansas State University, Manhattan, Kansas 66506, USA}
% ...
\affiliation{$^{65}$Michigan State University,
                East Lansing, Michigan 48824, USA}
% ...
\affiliation{$^{75}$University of Oklahoma, Norman, Oklahoma 73019, USA}
% ...
\date{September 2, 2008}

\begin{abstract}
A search for new physics in the acoplanar jet topology has been performed in 2.5\,\invfb\ 
of data from \ppb\ collisions at $\sqrt{s}=$1.96\,TeV, recorded by the D0 detector at
the Fermilab Tevatron Collider. 
The numbers of events with exactly two acoplanar jets and missing
transverse energy are in good agreement with the standard model
expectations.
The result of this search has been used to set a lower mass limit of 205\,GeV at the 95\% C.L.
on the mass of a scalar leptoquark when this particle decays exclusively into a quark and a neutrino.
In the framework of the Little Higgs model with T-parity, limits have also been obtained 
on the T-odd quark mass as a function of the T-odd photon mass.
\end{abstract}

\pacs{14.80.-j, 13.85.Rm}
\maketitle 

At hadron colliders, new colored particles predicted by various extensions of the standard
model (SM) would be abundantly produced if they are light enough.
The final state with jets and missing transverse energy ($\met$) resulting from the decay of those particles is a
promising channel to discover physics beyond the SM. In this Letter, a search for new
particles in the topology consisting of exactly two jets and $\met$ is presented using
2.5\,\invfb\ of data collected at a center-of-mass energy of 1.96\,TeV with the D0 detector
during Run~II of the Fermilab Tevatron \ppb\ Collider. The result of this search has been
used to constrain two categories of models.

The first category corresponds to models predicting the existence of leptoquarks (LQ)~\cite{lq}.
Those are scalar or vector particles carrying both a lepton and a baryon quantum number.
They are predicted by many extensions of the SM attempting to explain the apparent symmetry between
quarks and leptons. To satisfy experimental constraints on flavor changing neutral current interactions,
leptoquarks couple only within a single generation. Leptoquarks decay into a charged lepton and 
a quark with a branching ratio $\beta$, or into a neutrino and a quark with a branching ratio 
$1-\beta$. Pair production of leptoquarks assuming $\beta=0$ therefore leads only to a final state
consisting of two neutrinos and two quarks. 
The most stringent previous limit at 95\% C.L. on the scalar leptoquark mass of 136\,GeV~\cite{lqd0}
for $\beta=0$ was obtained by the D0 collaboration with 310\,\invpb\ of Run~II data.
The CDF collaboration also set a lower mass limit of 117\,GeV~\cite{lqcdf} with 191\,\invpb\ of Run~II data.
Those limits, as well as the results presented in this Letter, apply for first- and
second-generation scalar leptoquarks. For the third-generation, tighter limits were obtained
by increasing the signal sensitivity using heavy-flavor quark tagging~\cite{lq3d0}. 

The second category is the Little Higgs (LH) model~\cite{littlest}, which provides an interesting 
scenario for physics at
the TeV scale, predicting the existence of additional gauge bosons, fermions,  
and scalar particles with masses in the \mbox{100\,GeV -- 5\,TeV} range. 
Electroweak precision constraints are satisfied by introducing a discrete
symmetry called T-parity~\cite{LHT}.
This symmetry is constructed
such that all the SM states are even, while most new states of the LH model with T-parity (LHT)
are odd. In the LHT model, six new Dirac T-odd quarks (T-quarks or {$\tilde{Q}$) are the
partners of the left handed T-even quarks of the SM. In most of the parameter space, the 
lightest T-odd particle (LTP) is the  so-called ``heavy photon'' ($\tilde{A}_H$) which is stable
and weakly interacting. From SM precision measurements, it is possible to set a lower mass limit
of $\sim$80\,GeV on the mass of $\tilde{A}_H$~\cite{LTP}.
The new particle spectrum of the LHT model has similar properties to spectra of supersymmetric models. 
The LTP, just as the Lightest Supersymmetric Particle in SUSY models with R-parity conservation, is a
dark matter candidate which escapes undetected. There are, however, important differences: the new
T-odd particles have the same spin as their SM partner; and in the LHT model, 
some SM states, for example right-handed SM fermions or gluons, have no partners.
In the following, the mass of the T-quarks
from the first two generations is assumed to be degenerate, and pair production of those four T-quarks
is considered.
As the T-odd gauge bosons other than the $\tilde{A}_H$ are relatively heavy, 
T-quarks decay into a quark and $\tilde{A}_H$ in most of the parameter space accessible
at the Tevatron. It will be assumed in the following that this branching ratio is 100\%. 
Pair production of T-quarks therefore leads to a final state with two quarks and two LTP, giving the
missing transverse energy signature. The only direct constraint from collider data on the T-quark mass
is the $\sim$100\,GeV lower limit on the mass of the supersymmetric partner of the 
first two generations quarks from LEP~\cite{LEPsquark} which can also be applied to T-quarks. 
Prospective studies~\cite{mylh} have shown that the Tevatron can be
sensitive to T-quark masses up to $\sim$400\,GeV. This sensitivity is severely reduced
when the mass difference between the T-quarks and the LTP becomes small.

The D0 detector has been described in detail in Ref.~\cite{Abazov:2005pn}.
Tracks are reconstructed in a silicon microstrip tracker and a central fiber tracker (CFT),
both located within a 2~T superconducting solenoidal magnet. The liquid argon and uranium 
calorimeter consists of three cryostats. The central one covers
pseudorapidities~\cite{pseudorapidity}
$|\eta|\,\lsim\,1.1$, and the two end sections extend the coverage up to $|\eta|\,\approx\,4.2$.
The calorimeter is designed in projective towers of size $0.1\times 0.1$ in the $(\eta,\phi)$ plane,
where $\phi$ is the azimuthal angle in radians.
The outer muon system, covering $|\eta|\,<\,2$, consists of tracking detectors and scintillation 
trigger counters in front of 1.8~T iron toroids, followed by two similar layers after the toroids. 

Jets were reconstructed with the iterative midpoint cone algorithm~\cite{jetalgo} with cone radius
$\mathcal{R}= \sqrt{(\Delta\phi)^2+(\Delta y)^2}=0.5$ in azimuthal angle $\phi$ and rapidity
$y= \frac{1}{2} \ln [(E+p_{z})/(E-p_{z})]$.
The jet energy scale (JES) corrections were derived from the transverse momentum balance in photon-plus-jet 
events. The $\met$ was calculated from all calorimeter cells, and corrected for the jet energy 
scale and for the transverse momenta of reconstructed muons.

In events from SM processes, the presence of neutrinos from $W$ or $Z$ decay in the final state
generates large $\met$. The main irreducible SM background in this search for new particles is 
therefore the $Z(\to\nu\bar{\nu})$+jets process. The $W(\to l\nu)+$jets events also exhibit
the $\met$ signature, but their contribution can be significantly reduced by rejecting events
with an isolated electron or muon. However, the charged lepton can escape detection in 
uninstrumented regions of the detector, fail 
identification criteria, or be a tau lepton decaying hadronically. To further suppress that background,
events containing an isolated high $p_T$ track are rejected.
The other SM backgrounds for this search are the pair production of vector bosons
($WW$, $WZ$, $ZZ$) and the production of top quarks, either in pairs (\ttb) or via the electroweak
interaction.
Finally, multijet production when one or more jets are mismeasured also leads to a final state 
with jets and $\met$ (``QCD background'').

Events from SM processes and signal events were simulated using Monte Carlo (MC) generators and passed
through a full {\sc geant3}-based~\cite{geant} simulation of the detector geometry and response. They were
subsequently processed with the same reconstruction chain as the data. The parton distribution functions (PDFs)
used in the MC generators are the {\sc CTEQ6L1}~\cite{cteq6} PDFs. A data event from a randomly selected
beam crossing was overlaid on each event to simulate the additional minimum bias interactions
and detector noise. 
The {\sc alpgen} generator~\cite{Mangano:2002ea} was used to simulate $W/Z+$\,jets and $t\bar{t}$
production.
It was interfaced with {\sc pythia}~\cite{Sjostrand:2006za} for the simulation of initial
and final state radiation (ISR/FSR) and of jet hadronization. Pairs of vector bosons and electroweak top
quark production were simulated with {\sc pythia} and {\sc comphep}~\cite{Boos:2004kh}, respectively.
The next-to-leading order (NLO) cross sections were computed with {\sc mcfm\,5.1}~\cite{Campbell:2001ik}.
The QCD background was not simulated, since it can be conservatively neglected in the final stage 
of this analysis.

Leptoquark pair production and decays were simulated with {\sc pythia} and the {\sc CTEQ6L1} PDFs.
The LQ mass in the MC simulation ranged from 60 to 240\,GeV. The NLO cross sections of this process were
computed from a program based on~\cite{Kramer:1997hh} with a renormalization and factorization scale 
($\mu_{\mathrm{r,f}}$) equal to the LQ mass, and using the {\sc CTEQ6.1M} PDF sets.

For the LHT model, it has been shown in~\cite{mylh} that T-quark pair production and decay to 
$q\tilde{A}_H$ is very similar to squark pair production and decay to $q\xo$, where $\xo$
is the lightest neutralino. Signal efficiencies were therefore determined using MC events 
generated with {\sc pythia} corresponding to the production and decay of these supersymmetric
particles. It has been checked that the spin differences between the T-odd particles of the LHT model
and the supersymmetric particles do not modify the signal efficiencies. Therefore, MC simulations 
of such events were performed to cover the $\tilde{Q}$--$\tilde{A}_H$ mass plane accessible at the Tevatron.
Concerning the signal normalization, the cross section of first and second generation T-quark pair production is
equal to four times the cross section of heavy quark pair production, if no other new particles predicted by the LHT model
are involved in the T-quark production. The NLO cross sections of this signal were therefore calculated using
{\sc mcfm\,5.1}, with $\mu_{\mathrm{r,f}}$ equal to the T-quark mass, and the {\sc CTEQ6.1M} PDF sets.

\begin{table*}
\begin{center}
\begin{minipage}{\textwidth}
\caption{\label{lqcuts}
Number of events observed, expected from background and signal MC simulations, and signal efficiencies for $M_{LQ} = 200$\,GeV 
at the various stages of the analysis. The QCD multijet contribution is not included in the background contribution. The 
quoted uncertainties are the combined statistical and systematic uncertainties.
}
\begin{ruledtabular}
\begin{tabular}{lrD{;}{\,\pm\,}{-1}D{;}{\,\pm\,}{-1}D{;}{\,\pm\,}{-1}}
Cut applied & Data & \multicolumn{1}{c}{Background} & \multicolumn{1}{c}{Signal} & \multicolumn{1}{c}{Signal efficiency} \\
\hline
Preselection				 	      	      	& 208,055	& 30,752 ; 5350 & 166  ; 21   & 0.302 ; 0.037 \\
1st leading jet $p_T\,>\,35$\,GeV\footnotemark[1]          	& 122,456	& 25,352 ; 4410 & 152  ; 19   & 0.276 ; 0.034 \\
2nd leading jet $p_T\,>\,35$\,GeV\footnotemark[1]      		&  79,985	& 14,538 ; 2530 & 144  ; 18   & 0.262 ; 0.032 \\
$\met > $ 75~GeV					      	&   6,509	&  5,219 ;  909 & 125  ; 16   & 0.228 ; 0.028 \\
$\Delta\phi (\met,\mathrm{jet_1})\,>\,90^{\circ}$	      	&   6,386	&  5,148 ;  897 & 124  ; 15   & 0.226 ; 0.028 \\
$\Delta\phi_{\rm{min}}(\met,\mathrm{any\,jet})\,>\,50^{\circ}$  & 3,857 &  3,453 ;  602 &  93  ; 12   & 0.170 ; 0.021 \\
$\Delta\phi_{\rm{max}}(\met,\mathrm{any\,jet})\,<\,170^{\circ}$ & 2,855 &  2,568 ;  448 &  81  ; 10   & 0.147 ; 0.018 \\
Isolated electron veto			              		&   2,347 &  2,129 ;  371 &  79.1 ;  9.8  & 0.144 ; 0.018 \\
Isolated muon veto					      	&   2,007 &  1,880 ;  328 &  79.1 ;  9.8  & 0.144 ; 0.018 \\
Isolated track veto					      	&   1,472 &  1,398 ;  244 &  73.0 ;  9.1  & 0.133 ; 0.017 \\
Exactly two jets				      	      	&     957 &    858 ;  150 &  49.1 ;  6.1  & 0.089 ; 0.011 \\
Final $\HT$ cut				      	      		&  \multicolumn{4}{c}{optimized} \\
Final $\met$ cut				              	&  \multicolumn{4}{c}{optimized} \\
\end{tabular}
\footnotetext[1]{
First and second jets are also required to be central \mbox{($\etadet<0.8$)}, 
with an electromagnetic fraction below 0.95, and to have \mbox{$\rm{CPF0} \geq 0.75$}.
}
\end{ruledtabular}
\end{minipage}
\end{center}
\end{table*}

The analysis strategy follows closely the ``dijet'' analysis from Ref.~\cite{d0sqgl}.
Events were recorded using triggers requiring two acoplanar jets and large $\met$ or $\mht$, where
$\mht$ is the vector sum of the jet transverse momenta ($\mht = \vert\sum_{\mathrm{jets}} \overrightarrow{p_T}\vert$).
The trigger requirements evolved during the Run~II data taking period in order to take into account the increasing
peak instantaneous luminosity of the Tevatron. At the last stage of the trigger selection, the requirements
were typically the following: (1) $\met$ or $\mht$ greater than 30\,GeV and their separation from all jets greater
than $25^{\circ}$; (2) an azimuthal angle between the two highest $p_T$ jets less than $170^{\circ}$.
Offline, events where $\met$ was higher than 40\,GeV were then selected. The best primary vertex (PV0) was 
defined as the vertex with the smallest probability to be due to a minimum bias interaction~\cite{vtxreco}.
The longitudinal position of PV0 was required to be less than $60$\,cm from the 
detector center to ensure efficient vertex reconstruction. Good jets were defined as jets with a 
fraction of energy in the electromagnetic layers of the calorimeter lower than 0.95. The 
acoplanarity, i.e. the azimuthal angle between the two leading jets, $\mathrm{jet_1}$ and $\mathrm{jet_2}$,
ordered by decreasing transverse momentum, was required to be less than 165 degrees. 
Then, the two leading jets were required to be in the central region of the detector, 
with $\etadet\,<\,0.8$, where $\eta_{\mathrm{det}}$ is the jet pseudorapidity calculated under
the assumption that the jet originates from the detector center. After this preselection, 
the transverse momenta of the two leading jets had to be higher than 35 GeV. Finally, jets were required to originate from the best 
primary vertex, based on their associated tracks~\cite{d0sqgl}. This was accomplished by requiring
CPF0$\,>\,$0.75, where CPF0 is the fraction of track $p_T$ sum associated with the jet which comes from PV0,
CPF0$= \sum p^{{\rm track}}_T ({\rm PV0}) / \sum p^{{\rm track}}_T ({\rm any\ PV})$.

At this stage, the QCD multijet background is still largely dominant. To further reject those
events, the selection criteria on $\met$ was increased to 75 GeV.
The requirement that the azimuthal angle between the $\met$ and the first jet, 
$\Delta\phi (\met,\mathrm{jet_1})$, exceeds 90 degrees, was used to remove events where a jet was 
mismeasured and generating $\met$ aligned to that jet. Also, the minimal azimuthal angle 
$\Delta\phi_{\rm{min}}(\met,\mathrm{any\,jet})$ and the maximal azimuthal angle $\Delta\phi_{\rm{max}}(\met,\mathrm{any\,jet})$
between jets and $\met$ directions had to be greater than 50 degrees and lower than 170 degrees, respectively.

To suppress \mbox{$W(\rightarrow l\nu)$+jets} events, a veto on events containing an isolated electron 
or muon with \mbox{$p_T\,>\,$10\,GeV} was applied. 
Events with an isolated track were then rejected to further reduce that background.
Isolated tracks were required to have \mbox{$p_T\,>$\,5\,GeV}, to originate from PV0 with \mbox{$DCA(z)\,<$\,5\,cm}
and \mbox{$DCA(r)\,<$\,2\,cm},
where $DCA(z)$ and $DCA(r)$ are the positions of the projection of the distance of closest approach between 
the track and PV0 on the beam direction and in the plane transverse to the beamline, respectively.
The number of
hits in the CFT used to reconstruct the track was required to be at least 8. Finally, good quality tracks 
were selected by requiring the $\chi^2/dof$ of the track-fit reconstruction to be lower than 4.
A hollow cone with inner and outer radii of 0.06 and 0.5 was constructed around each track that passed those
criteria. If no other track with $p_T\,>$\,0.5\,GeV and the same quality criteria as above was found
in this hollow cone, the track was considered isolated. The use of a hollow, rather than full cone also
allowed rejection of tau leptons decaying into three charged particles.

Events with exactly two jets with $p_T\,>\,$15\,GeV and $|\eta_{det}|\,<\,$2.5 in the final state
were then selected. This criterion rejects a large fraction of the remaining $t\bar{t}$ events, and increases
the signal sensitivity at large T-quark and leptoquark masses once large $\met$ and $\HT$ are required,
with $\HT = \sum_{\mathrm{jets}} p_T$, where the sum is also over all jets with $p_T>15$\,GeV and $\etadet<2.5$.
Table~\ref{lqcuts} summarizes the number
of events observed and expected from MC simulations at each stage of the analysis. Figure~\ref{plots} shows comparisons
between data and MC simulations: the distribution of the number of jets, and the $\met$ and $\HT$ distributions
after applying all the selection criteria described above.

\begin{figure*}
\begin{tabular}{ccc}
\includegraphics[width=5.75cm]{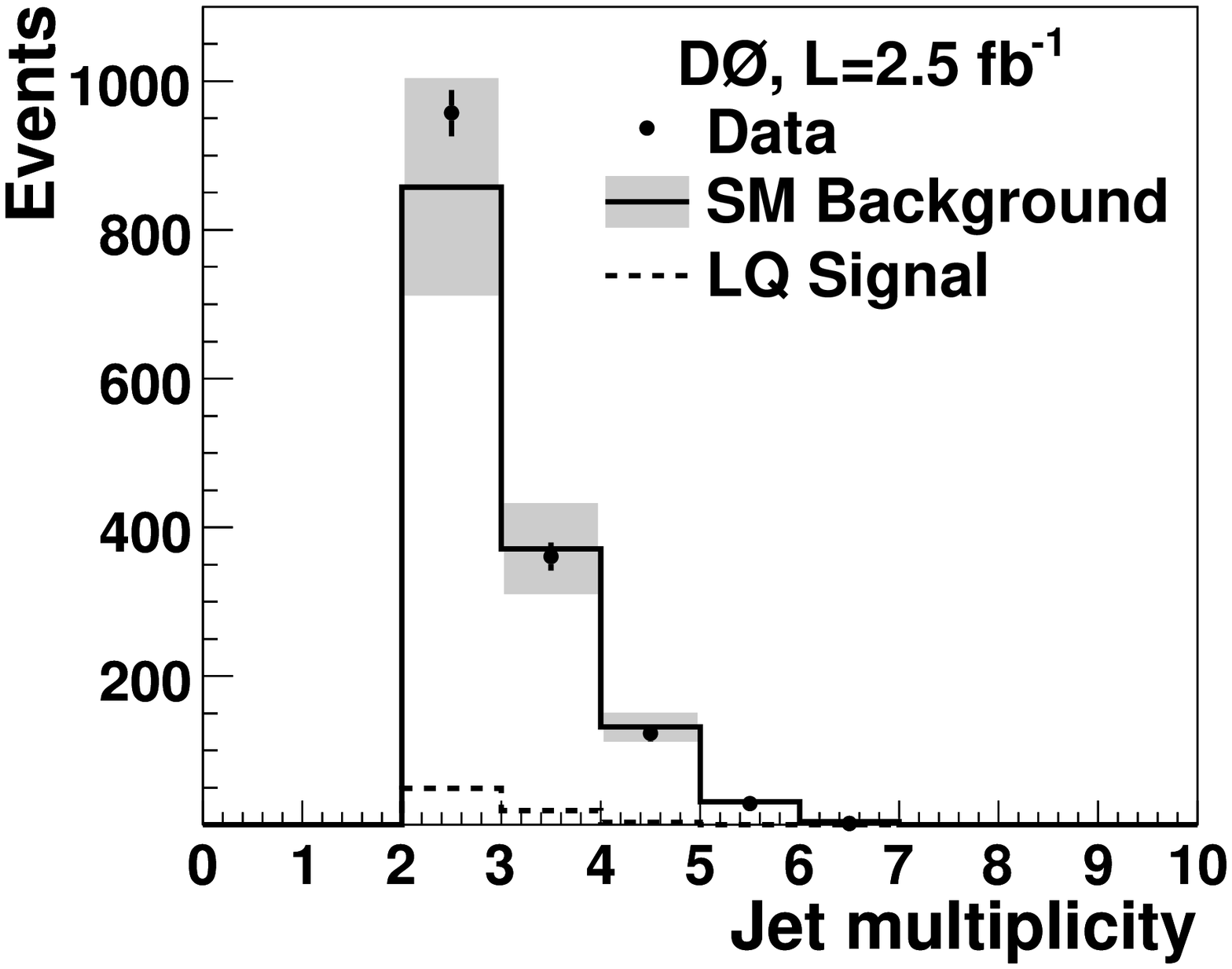} &
\includegraphics[width=5.75cm]{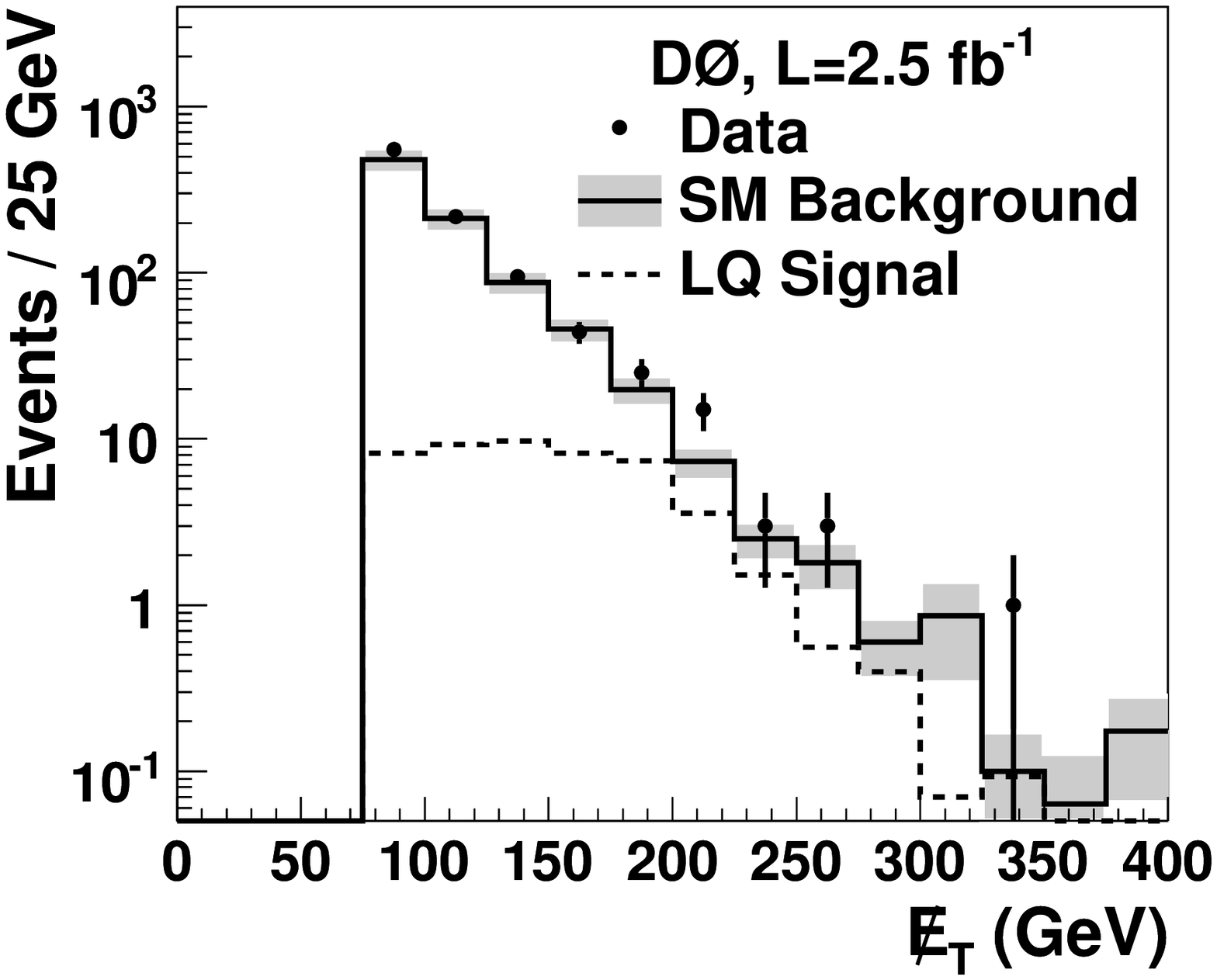} &
\includegraphics[width=5.75cm]{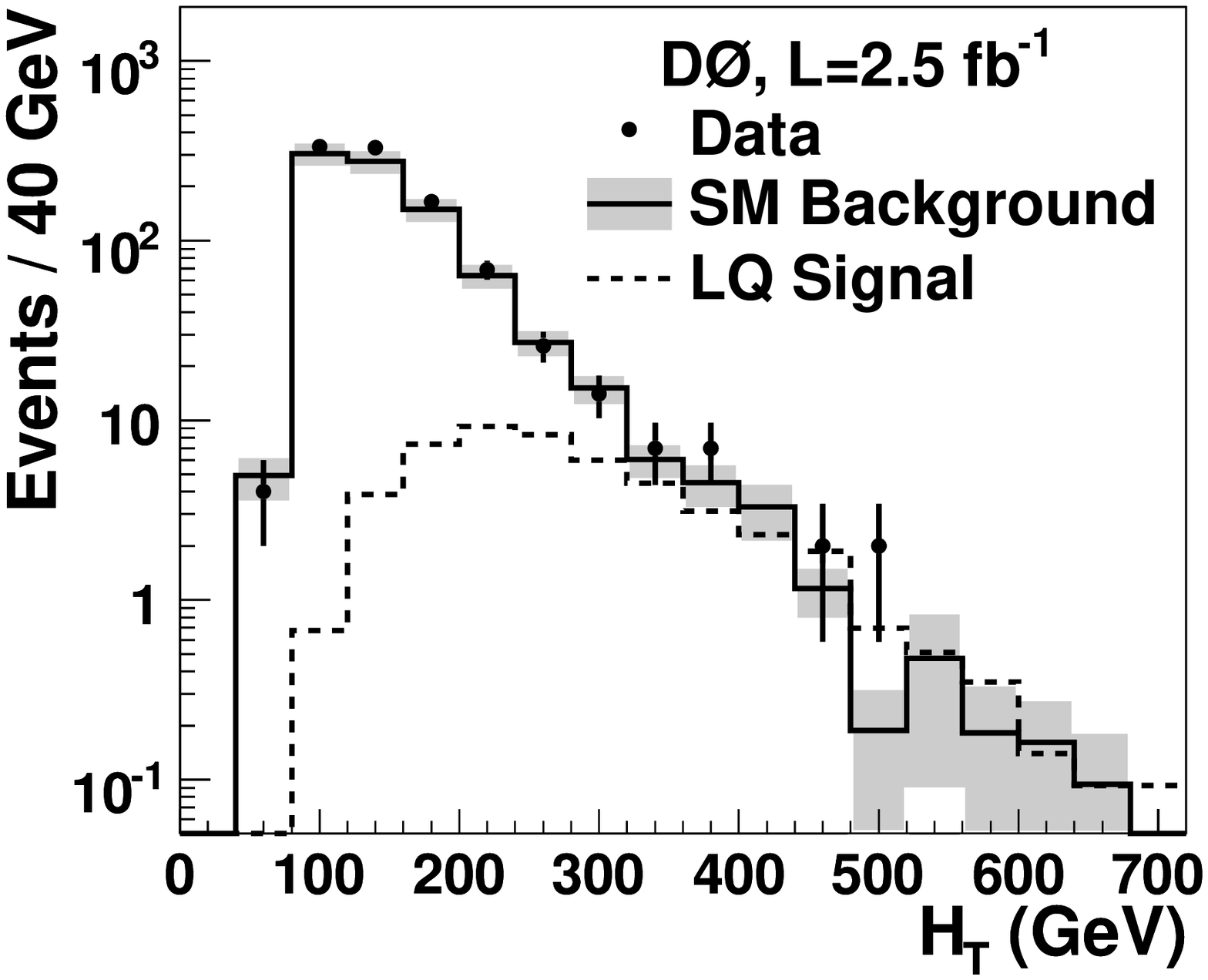} \\
(a) &
(b) &
(c) \\
\end{tabular}
\caption{\label{plots}
Distribution of the number of jets (a) and distributions of $\met$ (b) and $\HT$ (c) before the final
optimization of the cuts on these two quantities; for data (points with error bars), for SM background (full histogram
with shaded band corresponding to the total uncertainty), 
and for signal MC (dashed histogram). The signal drawn corresponds to pair production of scalar leptoquarks
with $M_{LQ}=200$\,GeV.
} 
\end{figure*}

Finally, the two final cuts on $\met$ and on $\HT$ were optimized for different signals by minimizing 
the expected upper limit on the cross section in the absence of signal. 
To this end and also for the final
limit computation, the CLs modified frequentist method has been used~\cite{CLS}.
For the leptoquark search,
two benchmarks were defined corresponding to low ($M_{LQ}=140$\,GeV) and high ($M_{LQ}=200$\,GeV) leptoquark
masses. As summarized in Table~\ref{summary1}, the optimized values were determined to be $\HT\,>\,150$\,GeV and $\met\,>\,75$\,GeV
for the low mass selection, and $\HT\,>\,300$\,GeV and $\met\,>\,125$\,GeV for the high mass selection. 
In the T-quark search, five $\HT$--$\met$ cut combinations were used to optimally scan the
($\tilde{Q},\tilde{A}_H$) mass plane as summarized
in Table~\ref{summary1}. In all cases, the contribution of the QCD multijet background was estimated to be small enough to be
conservatively neglected. The number of events observed are in good agreement with the SM expectations. 

\begin{table*}
\renewcommand{\arraystretch}{1.2}
\begin{center}
\caption{\label{summary1}
For each optimized event selection, information on the signal for which it was optimized 
($M_{LQ}$ or $(M_{\tilde{Q}},M_{\tilde{A}_H})$, and nominal NLO cross section), lower values of
$\HT$ and $\met$ selection criteria, the number of events observed, the number of events
expected from SM backgrounds, the number of events expected from signal, and the 95\% C.L.
signal cross section upper limit. The first uncertainty is statistical and the second one is 
systematic.
}
\begin{ruledtabular}
\begin{tabular}{ldcrD{;}{\,\pm\,}{-1}D{;}{\,\pm\,}{-1}d}
$M_{LQ}$ or $(M_{\tilde{Q}},M_{\tilde{A}_H})$		& \multicolumn{1}{c}{$\sigma_{\rm nom}$}	& $(\HT,\met)$ 	& $N_{\rm obs}$ & \multicolumn{1}{c}{$N_{\rm backgrd.}$}	& \multicolumn{1}{c}{$N_{\rm sig.}$} & \multicolumn{1}{c}{$\sigma_{95}$}\\
(GeV)							& \multicolumn{1}{c}{(pb)}			& (GeV)		&		& 						& & \multicolumn{1}{c}{(pb)} \\
\hline
\multicolumn{7}{c}{Leptoquark search} \\
\hline
140 							& 2.38			& $(150, 75)$	& 353		& 328  ;  11 ^{+56}_{-57} 			& 229  ; 8 ^{+24 }_{-23 } & 1.79  \\
200							& 0.268			& $(300,125)$	&  12		&  10.6 ;  1.7 ^{+4.0}_{-2.0} 			&  13.7 ; 0.6^{+ 1.8}_{- 2.0} & 0.240 \\
\hline
\multicolumn{7}{c}{T-quark search} \\
\hline
(150,100)						& 59.6			& $(125, 75)$	& 566		& 513  ; 14 ^{+86}_{-87} 			& 879  ; 167 ^{+108 }_{-94 } & 17.0    \\
(250,175)						&  3.18			& $(175,100)$	& 147		& 140  ; 7  ^{+25}_{-26} 			&  83  ;  12 ^{+ 16 }_{-10 } &  2.42   \\
(300,200)						&  0.868		& $(225,125)$	&  44		&  40  ; 4  ^{+ 7}_{- 7} 			&  25.7 ;   3.4^{+  4.3}_{- 4.7} &  0.780  \\
(350,200)						&  0.242		& $(275,150)$	&  15		&  13.1 ; 2.1 ^{+2.6}_{-2.7} 			&  16.4 ;   1.5^{+  3.1}_{- 3.0} &  0.169  \\
(400,150)						&  0.0666		& $(325,175)$	&   7		&   4.2 ; 1.0 ^{+1.2}_{-0.9}			&  10.1 ;   0.6^{+  1.2}_{- 1.5} &  0.0593 \\
\end{tabular}
\end{ruledtabular}
\end{center}
\end{table*}

The uncertainty coming from the JES corrections on the SM backgrounds and signal predictions
ranges from 5\% for low $\HT$ and $\met$ cuts to 10\% for high $\HT$ and $\met$ cuts.
The uncertainties due to the jet energy resolution, to the jet track confirmation, and to jet 
reconstruction and identification efficiencies range between 2\% and 4\%.
The systematic uncertainty due to the isolated track veto was measured to be 3\%.
All these uncertainties account for differences between data and MC simulation, both for signal
efficiencies and background contributions. The trigger was found to be fully efficient for the
event samples surviving all analysis requirements within an uncertainty of 2\%. 
The uncertainty on the luminosity measurement is 6.1\%~\cite{d0lumi}.
All of these uncertainties are fully correlated between signal and SM backgrounds. 
A 15\% systematic uncertainty was set on the $W/Z$+jets and \ttb\ NLO cross sections.
The uncertainty on the signal acceptance due to the PDF choice was 
determined to be 6\%, using the forty-eigenvector basis of the {\sc CTEQ6.1M} PDF set~\cite{cteq6}.
Finally, the effects of ISR/FSR on the signal efficiencies were studied by varying
the {\sc pythia} parameters controlling the QCD scales and the maximal allowed
virtualities used in the simulation of the space-like and time-like
parton showers. The uncertainty on the signal efficiencies was determined to be 6\%.

The nominal NLO signal cross sections, $\sigma_{\mathrm{nom}}$, were computed with the 
{\sc CTEQ6.1M} PDF and for the renormalization and factorization scale $\mu_{\mathrm{r,f}}=Q$, where $Q$ was taken to be equal 
to the leptoquark or T-quark mass. The uncertainty due to the choice of PDF was determined using the full set
of {\sc CTEQ6.1M} eigenvectors, with the individual uncertainties added in
quadrature. The effect of the renormalization and factorization scale was studied by calculating the signal cross sections
for $\mu_{\mathrm{r,f}}=Q$, $\mu_{\mathrm{r,f}}=Q/2$ and $\mu_{\mathrm{r,f}}=2 \times Q$.
The PDF and $\mu_{\mathrm{r,f}}$ effects were added in quadrature to compute minimum, 
$\sigma_{\mathrm{min}}$, and maximum, $\sigma_{\mathrm{max}}$, signal cross sections. 

For the leptoquark search, Fig.~\ref{exclulq} shows the 95\% C.L. observed and expected 
upper limits on scalar leptoquark production cross sections. The intersection
with the minimal NLO cross section gives a lower mass limit of 205\,GeV for $\beta=0$. 
The corresponding expected limit is 207\,GeV. Those limits are 214\,GeV and 222\,GeV,
respectively, for the nominal signal cross section.

For the T-quark search, Fig.~\ref{exclulht} shows the 95\% C.L. excluded regions in 
$\tilde{Q}$--$\tilde{A}_H$ mass plane assuming that the branching fraction of the decay
$\tilde{Q} \to q\tilde{A}_H$ is 100\%. The largest excluded T-quarks mass, 404\,GeV, is
obtained for large mass difference between the T-quarks and the LTP.

\begin{figure}
\includegraphics[width=8.5cm]{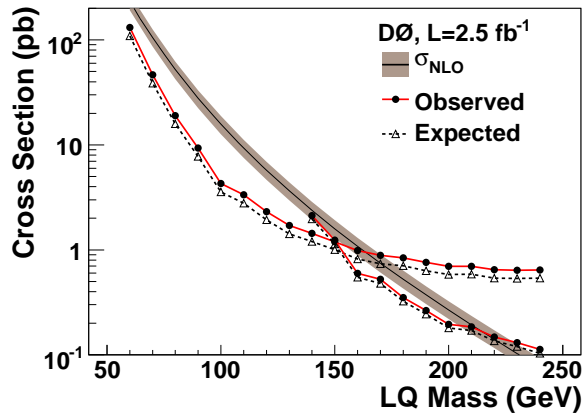}
\caption{\label{exclulq}
For the leptoquark search, observed (circles) and expected (triangles) 
95\% C.L. upper limits on scalar leptoquark production cross sections.
The limits obtained with the low mass and high mass selections are shown separately.
The nominal production cross sections are also shown for $\beta=0$, with shaded bands corresponding to the PDF and 
renormalization and factorization scale uncertainties.
}
\end{figure}

\begin{figure}
\includegraphics[width=8.5cm]{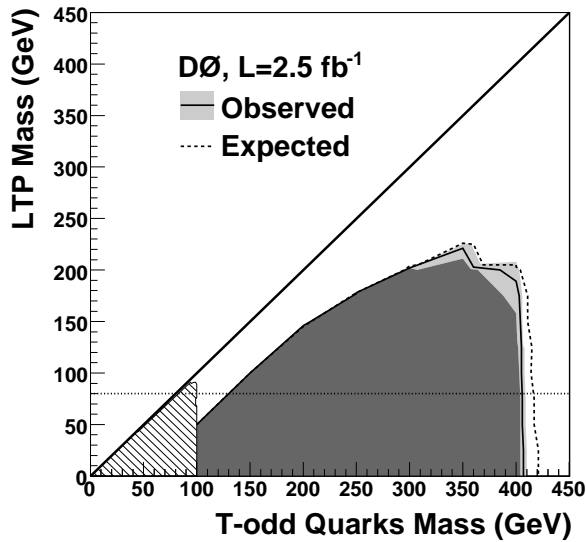}
\caption{\label{exclulht}
For the T-quark search, expected and observed 95\% C.L. excluded regions
in the $\tilde{Q}$--$\tilde{A}_H$ mass plane. The dark shaded region is the observed
exclusion for the minimal signal cross section hypothesis. The light shaded band shows
the effect on the observed exclusion coming from the theoretical uncertainty on the
signal cross section. The full and
dotted black lines are the observed and expected limits, respectively, for the nominal
cross section hypothesis.
The hatched region is excluded by LEP~\cite{LEPsquark}.
The region below the horizontal dashed line ($M_{\tilde{A}_H}\,<\,80$\,GeV) is
excluded by SM precision measurements~\cite{LTP}.
}
\end{figure}

In summary, a search for scalar leptoquarks and for T-quarks
produced in $p\bar{p}$ collisions at $\sqrt{s}=$1.96\,TeV has been performed with a 2.5\,\invfb\ data sample. 
This search was conducted in events containing exclusively two jets and large missing transverse energy.
The results are in good agreement with the SM background expectations, and 95\% C.L. limits have been
set on the leptoquark and T-quark masses. 
For a single-generation scalar leptoquark, a lower mass limit of 205\,GeV has been obtained for $\beta=0$, 
improving the previous
limit by 69\,GeV. In the LHT model,
limits on T-quark mass were obtained as a function of the $\tilde{A}_H$ mass assuming 100\% branching 
ratio for the decay $\tilde{Q} \to q\tilde{A}_H$. T-quark masses up to 404\,GeV are excluded when 
the mass difference between T-quarks and the LTP is large. Those are the most stringent direct limits to date
on the T-quarks mass.

%
We thank M.~Carena, J.~Hubisz, and M.~Perelstein for their
valuable help with the LHT model,
the staffs at Fermilab and collaborating institutions, 
and acknowledge support from the 
DOE and NSF (USA);
CEA and CNRS/IN2P3 (France);
FASI, Rosatom and RFBR (Russia);
CNPq, FAPERJ, FAPESP and FUNDUNESP (Brazil);
DAE and DST (India);
Colciencias (Colombia);
CONACyT (Mexico);
KRF and KOSEF (Korea);
CONICET and UBACyT (Argentina);
FOM (The Netherlands);
STFC (United Kingdom);
MSMT and GACR (Czech Republic);
CRC Program, CFI, NSERC and WestGrid Project (Canada);
BMBF and DFG (Germany);
SFI (Ireland);
The Swedish Research Council (Sweden);
CAS and CNSF (China);
and the
Alexander von Humboldt Foundation (Germany).
%


\begin{thebibliography}{99}
\bibitem{pseudorapidity}
The pseudorapidity $\eta$ is defined as
$-\ln \left[ \tan \left( \theta/2 \right) \right]$, with $\theta$ being the polar angle with respect
to the proton beam direction.
\end{thebibliography}
\end{document}